\documentclass[aip, jcp, amsmath, amssymb, reprint, longbibliography]{revtex4-2}
\usepackage{graphicx}
\usepackage{dcolumn}
\usepackage{bm}
\usepackage{color}
\usepackage{braket}
\usepackage{amsmath,amssymb,amsfonts,bm}

\begin{document}

\author{Gyanu P. Kafle}
\affiliation{Department of Chemistry, Wayne State University, Detroit, Michigan 48202, USA}
\author{Amos Afugu}
\affiliation{Department of Chemistry, Wayne State University, Detroit, Michigan 48202, USA}
\author{Zhen-Fei Liu}
\email{zfliu@wayne.edu}
\affiliation{Department of Chemistry, Wayne State University, Detroit, Michigan 48202, USA}
\title{Spin and Orbital Angular Momentum Textures at Chiral Molecule--Metal Interfaces}

\date{\today}

\begin{abstract}
Chiral molecular adsorption can modify the spin and orbital character of metal surface states, providing an electronic-structure basis for interpreting spin-dependent interfacial phenomena. Using first-principles calculations with spin--orbit coupling, band unfolding, and an atom-projected orbital angular momentum (OAM) analysis, we investigate spin and OAM textures at chiral heptahelicene/Au(111) interfaces and compare them with clean Au (111) and an achiral coronene/Au system. Adsorption changes the orientations of the spin and OAM vectors and reduces the sampled OAM magnitudes of Au-derived surface states near the Fermi level. The two heptahelicene enantiomers exhibit textures related by mirror and time-reversal symmetries, whereas the comparison with coronene identifies changes that also occur without molecular chirality. These results clarify how molecular adsorption modifies interfacial spin and orbital character and establish symmetry constraints relevant to chirality-dependent photoemission.
\end{abstract}

\maketitle

\section{Introduction}
Chiral molecule--metal interfaces are central to chirality-induced spin selectivity (CISS) \cite{bloom24} observed in photoemission \cite{Gohler2011,Kettner2018} and transport \cite{xie2011,albro2025} measurements. Understanding these phenomena requires examining how molecular handedness affects the electronic structure of the hybrid interface. Proposed mechanisms for spin selectivity involve orbital polarization and its conversion into spin polarization through spin--orbit coupling (SOC) \cite{Gersten2013,Adhikari2023}, which motivates the study of both spin and orbital angular momentum (OAM) textures, i.e., how spin and OAM are oriented in reciprocal space. Clean Au (111), a widely used nonmagnetic metal substrate in experiments, provides a useful reference, where the Shockley surface states near the Fermi level ($E_F$) exhibit Rashba-type spin splitting \cite{Lashell1996}. The latter arises from broken inversion symmetry at the surface and strong atomic SOC \cite{Henk2003,petersen2000}, while time-reversal symmetry constrains the resulting spin texture. Furthermore, the spin and OAM textures of the surface states have been characterized both experimentally \cite{Reniert2001,Nicolay2001,Hoesch2004} and theoretically \cite{Kim2012,Lee2012,Park2012,Kim2013}.

Although the spin and OAM textures of the clean Au (111) surface have been extensively studied, their modification upon adsorption of chiral molecules remains less explored. Molecular adsorption leads to hybridization between molecular and metal orbitals, resulting in modulation of the electronic structure as well as the spin and OAM textures near $E_F$. The spin and OAM textures at molecule--metal interfaces characterize the initial states in the photoemission process \cite{Hufner2003PES} from which a spin-dependent response can arise. For chiral molecules, a key question is which changes depend on molecular handedness and which also occur upon adsorption of an achiral molecule. The analysis of the spin and OAM textures therefore provides an electronic structure basis for interpreting CISS-related measurements.

In this work, we develop a computational framework, combining first-principles calculations, atom-projected OAM, and band unfolding at the interface, to investigate the spin and OAM textures at chiral molecule--Au(111) interfaces. Following our previous work on photoemission \cite{Afugu2026}, we consider two enantiomers, ($M$)-heptahelicene ($M$-[7]H, C$_{30}$H$_{18}$) and ($P$)-[7]H, which were shown to exhibit strong enantiomer-dependent photoelectron spin polarization \cite{Kettner2018}. Furthermore, we consider coronene (C$_{24}$H$_{12}$) as an achiral control system with a similar size to the [7]H, to identify changes in spin and OAM textures that also occur without molecular chirality. The present work complements our previous photoemission study at the $\Gamma$ point (normal emission) \cite{Afugu2026} by resolving the equilibrium spin and OAM textures around the Au-derived Fermi contours in the broader Brillouin zone.

This paper is organized as follows. Sec. \ref{sec:methods} describes the computational framework that we have developed to compute the spin and OAM textures from first-principles calculations of periodic interfaces. We first apply these approaches to the well-studied clean Au (111) surface and show the results in Sec. \ref{sec:auuc}, validating our approach. Then in Sec. \ref{sec:7hau}, we apply our approach to ($M$)- and ($P$)-[7]H/Au interfaces, focusing on the spin and OAM characters of the spin-split Shockley surface states. The spin and OAM textures of these interfaces are reported in Sec. \ref{sec:texture} and compared to those of the clean Au (111) and the achiral coronene/Au interface. We conclude in Sec. \ref{sec:conclude} and discuss how our results provide insight into CISS observed in photoemission experiments.

\section{Methods}
\label{sec:methods}

\subsection{Spin expectation values}

We evaluate spin and OAM expectation values from normalized two-component Kohn--Sham spinor wave functions from density functional theory (DFT),
\begin{equation}
\ket{\psi_{n\mathbf{k}}} =
\ket{\phi_{1,n\mathbf{k}}}\ket{\uparrow}+
\ket{\phi_{2,n\mathbf{k}}}\ket{\downarrow},
\label{eq:spinor}
\end{equation}
where $n$ is the band index, $\mathbf{k}$ is the crystal momentum, and $\ket{\uparrow}$ and $\ket{\downarrow}$ are simultaneous eigenstates of $\hat{S}^2$ and $\hat{S}_z$. $\ket{\phi_{1,n\mathbf{k}}}$ and $\ket{\phi_{2,n\mathbf{k}}}$ represent the spatial wave functions associated with the spin-up and spin-down components, respectively. 

For $\ket{\psi_{n\mathbf{k}}}$, the spin expectation values are:
\begin{subequations}
\begin{align}
\braket{S_x}_{n\mathbf{k}}
&= \hbar\,\mbox{Re}\braket{\phi_{1,n\mathbf{k}} | \phi_{2,n\mathbf{k}}}, \\
\braket{S_y}_{n\mathbf{k}}
&= \hbar\,\mbox{Im}\braket{\phi_{1,n\mathbf{k}} | \phi_{2,n\mathbf{k}}}, \\
\braket{S_z}_{n\mathbf{k}}
&= \frac{\hbar}{2}
\left(\braket{\phi_{1,n\mathbf{k}} | \phi_{1,n\mathbf{k}}}-\braket{ \phi_{2,n\mathbf{k}} | \phi_{2,n\mathbf{k}}}\right).
\end{align}
\label{eq:sxyz}
\end{subequations}
Here, $\hat{S}_\alpha$ with $\alpha=x,y,$ and $z$ are spin operators, i.e., $S_\alpha=\hbar/2\sigma_\alpha$ with $\sigma_\alpha$ the Pauli matrices.

\subsection{On-site OAM character with SOC}

We characterize OAM using its on-site component, for which we develop an atom-projected approach in evaluating the OAM expectation values. We first project Eq. \eqref{eq:spinor} onto the L\"{o}wdin-orthogonalized atom-centered basis functions, using subroutines provided by Quantum ESPRESSO \cite{QE}:
\begin{equation}
\ket{\psi_{n\mathbf{k}}}
=\sum_{\mu}\sum_{\ell,j,m_j} c^{\ell,j,m_j;\mu}_{n\mathbf{k}} \ket{\ell,j,m_j;\mu},
\label{eq:exp}
\end{equation}
where the sums run over atom index $\mu$, OAM quantum number $\ell$, total angular momentum quantum number $j$, and total angular momentum magnetic quantum number $m_j$. Because $\ket{\ell,j,m_j;\mu}$ is not an eigenvector of the OAM operator $\hat{L}_\alpha$, the coupled basis $\ket{\ell,j,m_j;\mu}$ is then transformed to the uncoupled basis $\ket{\ell,m_\ell;\mu}\ket{\uparrow}$ and $\ket{\ell,m_\ell;\mu}\ket{\downarrow}$ via the Clebsch--Gordan coefficients. Here, $\ket{\ell,m_\ell;\mu}$ are atom-centered spherical harmonics and simultaneous eigenstates of $\hat{L}^2$ and $\hat{L}_z$, where $m_\ell$ is the orbital magnetic quantum number. 

As implemented in Quantum ESPRESSO, for $j=\ell+1/2$, the above transformation is:
\begin{equation}
\begin{aligned}
\ket{\ell,\ell+\frac{1}{2},m_j;\mu}& =\sqrt{\frac{\ell+m_j+\frac{1}{2}}{2\ell+1}}
\ket{\ell,m_j-\frac{1}{2};\mu}\ket{\uparrow} \\
& +\sqrt{\frac{\ell-m_j+\frac{1}{2}}{2\ell+1}} \ket{\ell,m_j+\frac{1}{2};\mu}\ket{\downarrow}.
\end{aligned}
\label{eq:cg1}
\end{equation}
For $j=\ell-1/2$, this is:
\begin{equation}
\begin{aligned}
\ket{\ell,\ell-\frac{1}{2},m_j;\mu} & =\sqrt{\frac{\ell-m_j+\frac{1}{2}}{2\ell+1}}
\ket{\ell,m_j-\frac{1}{2};\mu}\ket{\uparrow} \\
& -\sqrt{\frac{\ell+m_j+\frac{1}{2}}{2\ell+1}} \ket{\ell,m_j+\frac{1}{2};\mu}\ket{\downarrow}.
\end{aligned}
\label{eq:cg2}
\end{equation}
Combining Eqs. \eqref{eq:exp}--\eqref{eq:cg2}, we can write:
\begin{equation}
\ket{\psi_{n\mathbf{k}}}= \sum_{\ell,m_\ell;\mu}\left(P_{n\mathbf{k}}^{\ell,m_\ell;\mu}\ket{\uparrow}+Q_{n\mathbf{k}}^{\ell,m_\ell;\mu}\ket{\downarrow}\right)\ket{\ell,m_\ell;\mu}.
\label{eq:lmlexp}
\end{equation}
Here, we denote $P$ and $Q$ as the \emph{normalized} expansion coefficients for the spin-up and spin-down channels, respectively. They can be obtained by multiplying $c^{\ell,j,m_j;\mu}_{n\mathbf{k}}$ in Eq. \eqref{eq:exp} by the Clebsch--Gordan coefficients in Eqs. \eqref{eq:cg1} and \eqref{eq:cg2} followed by a common normalization over the two spin channels. We note that Eq. \eqref{eq:lmlexp} may not be a complete expansion due to the incompleteness of the $\ket{\ell,m_\ell;\mu}$ basis. Therefore, we normalize Eq. \eqref{eq:lmlexp} before evaluating $\braket{L_\alpha}$.

The strategy in computing $\braket{L_\alpha}_{n\mathbf{k}}$ is the following: because $\hat{L}_\alpha$ does not act on spin,  $\braket{L_\alpha}$ can be evaluated first for the spin-up and spin-down components in Eq. \eqref{eq:lmlexp} separately, before being summed up over the spin channel. 

For $\braket{L_z}$, because $\ket{\ell,m_\ell;\mu}$ is an eigenket of $\hat{L}_z$, i.e., $\hat{L}_z \ket{\ell,m_\ell;\mu}=\hbar m_\ell \ket{\ell,m_\ell;\mu}$, we conclude
\begin{equation}
\braket{L_z}_{n\mathbf{k}}= \sum_{\ell,m_\ell;\mu}\hbar m_l\left(\left|P_{n\mathbf{k}}^{\ell,m_\ell;\mu}\right|^2+\left|Q_{n\mathbf{k}}^{\ell,m_\ell;\mu}\right|^2\right).
\label{eq:lz}
\end{equation}

To evaluate $\braket{L_x}$ and $\braket{L_y}$, we need the ladder operator, $\hat{L}_{+}=\hat{L}_x+i\hat{L}_y$, and the relation $\hat{L}_{+}\ket{\ell,m_\ell;\mu}=\hbar\sqrt{\ell(\ell+1)-m_\ell(m_\ell+1)}\ket{\ell,m_\ell+1;\mu}$. Using Eq. \eqref{eq:lmlexp}, we first compute
\begin{equation}
\begin{aligned}
\braket{L_{+}}_{n\mathbf{k}} & = \sum_{\ell,m_\ell;\mu} \hbar \sqrt{\ell(\ell+1)-m_\ell(m_\ell+1)} \\
& \times \left[\left(P_{n\mathbf{k}}^{\ell,m_\ell+1;\mu}\right)^*P_{n\mathbf{k}}^{\ell,m_\ell;\mu}+ \left(Q_{n\mathbf{k}}^{\ell,m_\ell+1;\mu}\right)^*Q_{n\mathbf{k}}^{\ell,m_\ell;\mu}\right]. \\
\end{aligned}
\label{eq:lplusexp}
\end{equation}
Then we have
\begin{subequations}
\begin{align}
\braket{L_x}_{n\mathbf{k}} & =\mbox{Re}\braket{L_{+}}_{n\mathbf{k}},\\
\braket{L_y}_{n\mathbf{k}} & =\mbox{Im}\braket{L_{+}}_{n\mathbf{k}}.
\end{align}
\label{eq:lxy}
\end{subequations}

We note that the above atom-projected approach measures the local OAM character associated with each $\ket{\psi_{n\mathbf{k}}}$, which neglects the interstitial and itinerant contributions to $\braket{L_\alpha}$. Formally, one needs to employ the modern theory of orbital magnetization \cite{Thonhauser2005, Lopez2012} for a gauge-invariant description of the full orbital circulation of an extended Bloch state. However, the atom-projected approach developed here provides a physically transparent and computationally efficient descriptor of the local, atom-resolved OAM texture. This is particularly useful to elucidate how molecular adsorption and molecule--metal hybridization modify the atomic-orbital composition and local OAM character, which is the central goal of this work. 

In this section, we have retained $\hbar$ in the equations to make the units explicit. In what follows, numerical spin and OAM values are expressed in units of $\hbar$.

\subsection{Band unfolding from the supercell}
\label{sec:unfolding}

Although the Au (111) surface states can be easily identified from band structure calculations of the unit cell, the molecule--metal interfaces involve supercells, which complicates their identification in the band structure due to band folding. To resolve this issue, we perform band unfolding from the supercell into the unit cell. Physically, we need to determine how much a supercell band $\ket{n\mathbf{k}_{\rm sc}}$ contains Bloch-state character of the primitive unit cell with crystal momentum $\mathbf{k}_{\rm uc}$. In general, $\mathbf{k}_{\rm uc}=\mathbf{k}_{\rm sc}+\mathbf{G}_{0,{\rm sc}}$, i.e., they differ by a supercell reciprocal-space lattice vector.

We follow the strategy proposed by Choi as implemented in Ref. \citenum{Choi_qe_unfolding}, which is consistent with established approaches in the literature \cite{Dargam1997, Wang1998, Boykin2007, Dirinberger2021, Quan2026}. Specifically, for our purpose of identifying Au (111) surface states and the related spin and OAM textures at $E_F$, the above problem can be simplified by considering $\mathbf{k}_{\rm uc}=\mathbf{k}_{\rm sc}$. This is because the target $\mathbf{k}_{\rm uc}$ -- the Fermi wave vector of Au (111) surface states -- is approximately 0.2 \AA$^{-1}$ (see, e.g., Ref. \citenum{Lashell1996}), which comfortably lies within the first Brillouin zone of a $4\times 4$ Au (111) supercell. Therefore, we only need to consider $\mathbf{k}_{\rm uc}$ values in the much smaller region of the \emph{supercell} Brillouin zone and do not distinguish $\mathbf{k}_{\rm uc}$ and $\mathbf{k}_{\rm sc}$ in the equations below. 

A normalized supercell Bloch state can be written as
\begin{equation}
    \psi_{n\mathbf{k}}(\mathbf{r})=e^{i\mathbf{k}\cdot \mathbf{r}}\sum_{\mathbf{G}_{\rm sc}}C_{n\mathbf{k}}\left(\mathbf{G}_{\rm sc}\right)e^{i\mathbf{G}_{\rm sc}\cdot \mathbf{r}},
    \label{eq:scpsi}
\end{equation}
where $\mathbf{G}_{\rm sc}$ runs over the reciprocal-space lattice vectors of the supercell. Determining its unit-cell character with crystal momentum $\mathbf{k}$ is equivalent to selecting the plane-wave components where $\mathbf{G}_{\rm sc}$ is also a $\mathbf{G}_{\rm uc}$. Formally, this can be done by defining a projection operator
\begin{equation}
\hat{P}_{\rm uc}=\sum_{\mathbf{G}_{\rm uc}}\ket{\mathbf{k}+\mathbf{G}_{\rm uc}}\bra{\mathbf{k}+\mathbf{G}_{\rm uc}},
\label{eq:proj}
\end{equation}
and assigning
\begin{equation}
W_{n\mathbf{k}} =\braket{\psi_{n\mathbf{k}}|\hat{P}_{\rm uc}|\psi_{n\mathbf{k}}}=\sum_{\mathbf{G}_{\rm sc}\in\cal{G}} \left|C_{n\mathbf{k}}\left(\mathbf{G}_{\rm sc}\right)\right|^2
\label{eq:weight}
\end{equation}
as the weight or unit-cell character (with unit-cell crystal momentum $\mathbf{k}$) associated with a supercell band $\ket{n\mathbf{k}}$. $\cal{G}$ is defined as the set of $\mathbf{G}_{\rm sc}$ that is also a $\mathbf{G}_{\rm uc}$. In practice, $\cal{G}$ can be found by the following approach: for a specific $\mathbf{G}_{\rm sc}$ under consideration, we write $\mathbf{G}_{\rm sc}=\mathbf{g}_{\rm sc}\mathbf{B}_{\rm sc}$, where $\mathbf{g}_{\rm sc}$ is an integer row vector and the rows of $\mathbf{B}_{\rm sc}$ are reciprocal-lattice vectors. If we write $\mathbf{A}_{\rm sc}=S\mathbf{A}_{\rm uc}$, with real-space lattice vectors stored as rows, then $\mathbf{B}_{\rm sc}=(S^{-1})^T\mathbf{B}_{\rm uc}$. Thus, $\mathbf{G}_{\rm sc}\in\mathcal{G}$ if and only if every component of $\mathbf{g}_{\rm sc}(S^{-1})^T$ is an integer. 

\section{Results and Discussion}

\subsection{Clean Au (111) surface}
\label{sec:auuc}

\begin{figure}[htb]
\centering
\includegraphics[width=\columnwidth]{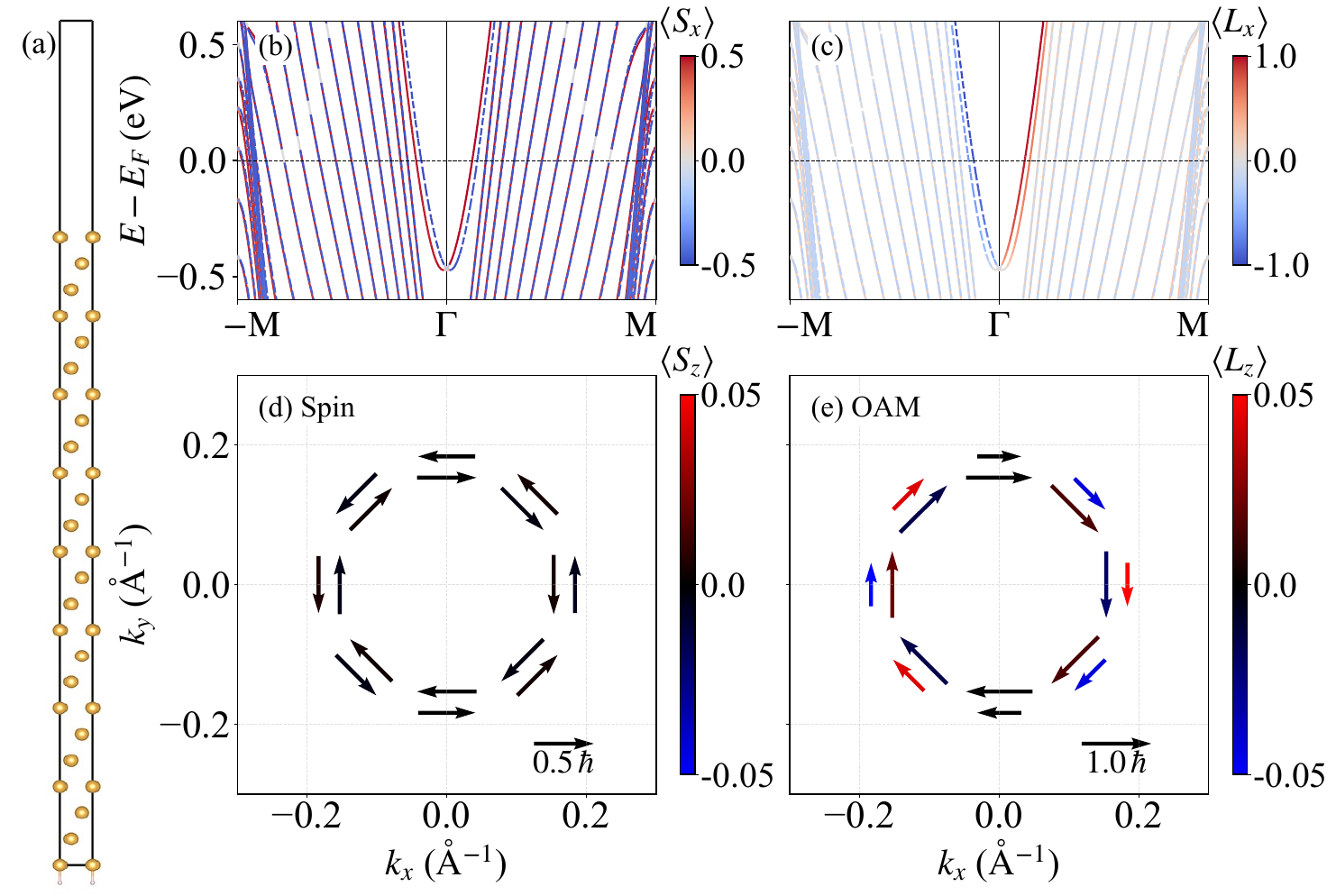}
\caption{(a) Atomic structure of a 25-layer Au (111) slab in its unit cell, with hydrogen passivation at the bottom surface. Electronic band structures with each state $\ket{n\mathbf{k}}$ color-coded by (b) $\braket{S_x}_{n\mathbf{k}}$ and (c) $\braket{L_x}_{n\mathbf{k}}$. (d) Spin and (e) OAM texture at $E_F$ near the $\Gamma$ point, highlighting the Rashba-split surface states. The out-of-plane components, $\braket{S_z}$ and $\braket{L_z}$, are denoted by color.}
\label{fig1:cartoon}
\end{figure}

We first examine a well-studied system, the clean Au (111) surface, to validate our computational approach and provide a baseline for the molecule--metal interfaces studied below. We model the clean Au (111) surface using a 25-layer unrelaxed slab, with the in-plane Au--Au distance being 2.885 \AA~(corresponding to the experimental face-centered cubic lattice constant of 4.08 \AA). In principle, one needs a sufficiently thick Au (111) slab \cite{Yan2015} to effectively reduce the coupling between its top and bottom surfaces. Alternatively, we found that an effective way to decouple the two surfaces is to passivate one surface with hydrogen atoms. As a result, the energies of the Shockley surface states can be converged with a much smaller number of Au layers. The structure is shown in Fig. \ref{fig1:cartoon}(a), where the bottom surface is passivated with hydrogen atoms, with an Au--H bond length of 1.6 \AA.

We compute its electronic structure using the Perdew--Burke--Ernzerhof (PBE) functional \cite{PBE} with SOC as implemented in Quantum ESPRESSO \cite{QE}, the optimized norm-conserving Vanderbilt (ONCV) pseudopotentials \cite{norn-conserving}, a $16\times 16\times 1$ $\mathbf{k}$-point mesh, and a 70 Ry kinetic energy cutoff. The electronic band structure is shown in Fig. \ref{fig1:cartoon}(b)(c), along the $k_y$ axis, corresponding to the ($-$M)--$\Gamma$--M line [see Fig. \ref{fig2:struc}(d) for the Brillouin zone]. The bands are color-coded by $\braket{S_x}$ in (b) and by $\braket{L_x}$ in (c). The characteristic Rashba spin-split surface states can be clearly identified near $E_F$, but their $\braket{S_x}$ and $\braket{L_x}$ values exhibit a distinct difference: at a fixed non-zero $k_y$, the two Rashba-split surface-state branches have opposite signs of $\braket{S_x}$ but the same sign of $\braket{L_x}$. Reversing $k_y$ reverses the signs of both spin and OAM.

The above distinction is further clearly illustrated in Fig. \ref{fig1:cartoon}(d)(e), where we plot the spin and OAM textures, evaluated around the $\Gamma$ point in the Brillouin zone at $E_F$ (to be precise, numerically within a narrow energy window $E_F\pm1$ meV), highlighting the Rashba-split surface states. Each arrow is centered at a sampled $(k_x,k_y)$ point where a surface state falls within the narrow energy window. Its in-plane components are $(\braket{S_x},\braket{S_y})$ in Fig. \ref{fig1:cartoon}(d) or $(\braket{L_x},\braket{L_y})$ in Fig. \ref{fig1:cartoon}(e). Its color represents the corresponding out-of-plane component. The much narrower color scale in Fig. \ref{fig1:cartoon}(d)(e) resolves out-of-plane components that are much smaller than the dominant in-plane components.

For the spin texture, $\braket{S_z}$ is nearly zero in the $(k_x,k_y)$ plane, with the inner and outer contours having opposite helicities but similar magnitudes. In contrast, for the OAM texture, the inner contour features larger magnitudes than the outer contour, and the arrows are oriented in the same direction. These qualitative features are consistent with the literature \cite{Henk2003, Hoesch2004, Kim2012, Lee2012}. 

Furthermore, our calculations reveal small but non-zero $\braket{S_z}$ and $\braket{L_z}$ for the surface states at certain $(k_x,k_y)$ points, with magnitudes ranging between 0 and 0.01 $\hbar$ for $\braket{S_z}$ and between 0 and 0.05 $\hbar$ for $\braket{L_z}$. Although previous calculations\cite{Kim2012, Lee2012} did not explicitly report these out-of-plane $\braket{S_z}$ and $\braket{L_z}$, a non-zero out-of-plane component for axial vectors (spin and OAM) is symmetry-allowed at a generic $(k_x,k_y)$ \cite{Henk2004}, even though it must vanish on a mirror-invariant line for a non-degenerate state. For the Au (111) slab used here, $yz$ is a mirror plane, which means that only the $x$-component of an axial vector is allowed on the $k_y$ axis. Our results strictly satisfy this symmetry constraint.

Among all the $\mathbf{k}$ points we sampled, the largest $\left|\braket{\mathbf{S}}\right|$ is $0.50\hbar$, while the largest $\left|\braket{\mathbf{L}}\right|$ is $0.95\hbar$. Here, $\left|\braket{\mathbf{S}}\right|=\sqrt{\braket{S_x}^2+\braket{S_y}^2+\braket{S_z}^2}$ and $\left|\braket{\mathbf{L}}\right|$ is similarly defined. Both the normalization of Eq. \eqref{eq:lmlexp} and the number of Au layers affect the calculated $\left|\braket{\mathbf{L}}\right|$ value. 

\subsection{Electronic band structures of [7]H/Au interfaces}
\label{sec:7hau}

\begin{figure}[!ht]
\centering
\includegraphics[width=\columnwidth]{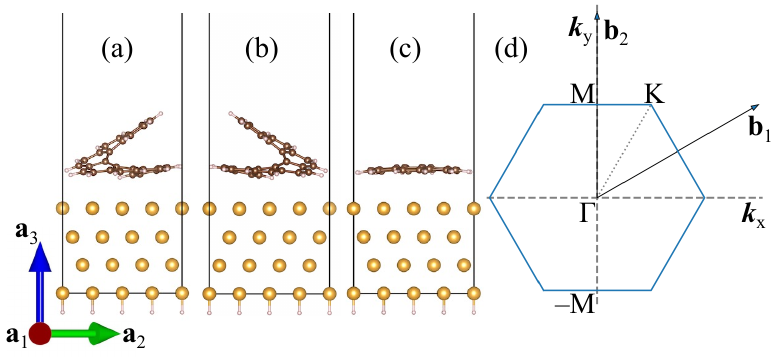}
\caption{Atomic structures of interfaces studied in this work: (a) ($M$)-[7]H, (b) ($P$)-[7]H, and (c) coronene adsorbed on Au (111). The bottom Au layer is passivated with hydrogen atoms. (d) 2D hexagonal Brillouin zone, applicable to all systems considered in this work, showing reciprocal-space lattice vectors $\mathbf{b}_1$ and $\mathbf{b}_2$, as well as high-symmetry $\mathbf{k}$ points $\Gamma$, M, and K.}
\label{fig2:struc}
\end{figure}

The three molecule--metal interfaces are shown in Fig. \ref{fig2:struc}: the chiral ($M$)-[7]H/Au(111) interface is shown in Fig. \ref{fig2:struc}(a), the chiral ($P$)-[7]H/Au(111) interface in Fig. \ref{fig2:struc}(b), and the achiral control, coronene/Au(111), in Fig. \ref{fig2:struc}(c). To remove differences between independently optimized adsorption geometries and focus on the electronic structure, the ($P$)-[7]H/Au(111) interface structure was constructed as the exact reflection of every atom (including Au and the passivating hydrogen) of the ($M$)-[7]H/Au(111) structure through the $xz$ plane. 

All systems contain a $4\times4$ Au (111) supercell consisting of four Au atomic layers, with the bottom two layers fixed during geometry relaxation, in which we employed the vdw-DF-cx functional \cite{vdw-df-cx} and similar computational parameters as in our previous work\cite{Afugu2026}. As in the Au (111) unit cell discussed in Sec. \ref{sec:auuc}, the bottom Au atoms are passivated by hydrogen atoms. We have checked that, with the hydrogen passivation, the energies of the Au (111) surface band are nearly converged for a four-layer Au (111) slab. To be explicit, we monitor the energy of the surface state at the $\Gamma$ point for perfect Au (111) slabs with varying numbers of Au atomic layers. This energy is 0.49 eV below $E_F$ for a four-layer slab and 0.47 eV for a nine-layer slab (the system used in our previous work\cite{Afugu2026}), compared to 0.47 eV for the 25-layer slab reported above. The convergence of this energy, together with that of the surface band dispersion, suggests the quantitative accuracy of the Fermi wave vectors of the surface states in our calculations.

\begin{figure}[!ht]
\centering
\includegraphics[width=\columnwidth]{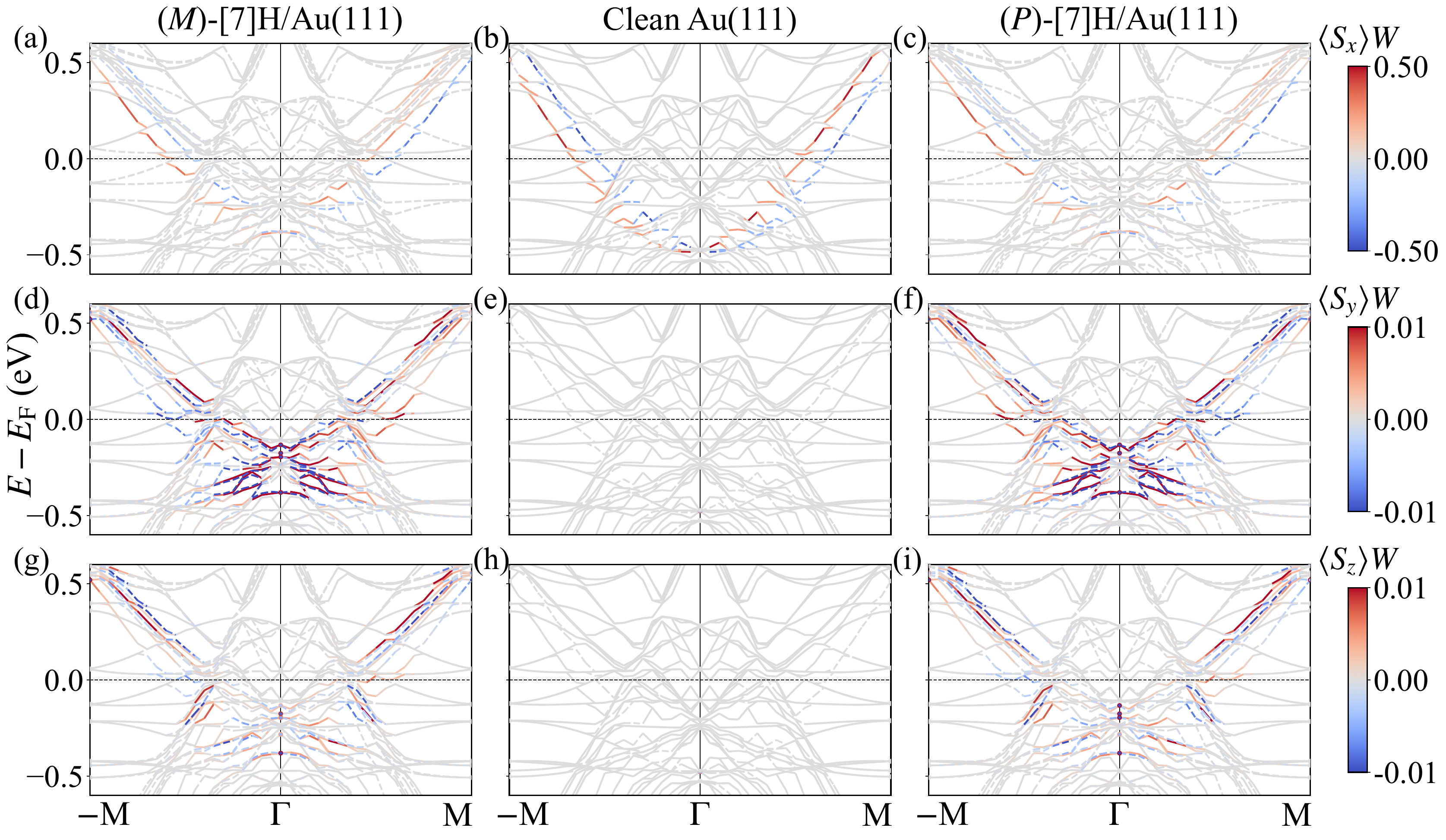}
\caption{Electronic band structures of (a)(d)(g) ($M$)-[7]H/Au(111) interface, (b)(e)(h) clean Au (111) surface, and (c)(f)(i) ($P$)-[7]H/Au(111) interface, referenced to $E_F$ of each system. To highlight the metal surface states, the bands are color-coded by the spin expectation values [Eq. \eqref{eq:sxyz}] multiplied by the unit-cell character [Eq. \eqref{eq:weight}]: (a)(b)(c) $\braket{S_x}_{n\mathbf{k}}W_{n\mathbf{k}}$, (d)(e)(f) $\braket{S_y}_{n\mathbf{k}}W_{n\mathbf{k}}$, and (g)(h)(i) $\braket{S_z}_{n\mathbf{k}}W_{n\mathbf{k}}$. Note that the color scales are much narrower for $\braket{S_y}$ and $\braket{S_z}$.}
\label{fig3:7H-Au-MGM}
\end{figure}

We used similar computational parameters as for the Au (111) unit cell in Sec. \ref{sec:auuc}, except that the $\mathbf{k}$-point mesh is now $4\times 4 \times 1$. Fig. \ref{fig3:7H-Au-MGM} shows the electronic band structures of ($M$)-[7]H/Au(111) (left three panels) and ($P$)-[7]H/Au(111) (right three panels), plotted along the ($-$M)--$\Gamma$--M direction along the $k_y$ axis [see Fig. \ref{fig2:struc}(d) for the Brillouin zone]. To illustrate the molecular effects, we compare them to those of the clean Au (111) surface, also modeled in a $4\times4$ supercell with four Au atomic layers. The latter is a perfect, unrelaxed Au (111) slab with hydrogen passivation at the bottom layer, whose band structure is shown in the middle three panels of Fig. \ref{fig3:7H-Au-MGM}. The band structures are color-coded by the spin expectation values [Eq. \eqref{eq:sxyz}] multiplied by the unit-cell character [Eq. \eqref{eq:weight}]: $\braket{S_x}_{n\mathbf{k}}W_{n\mathbf{k}}$ in the top three panels, $\braket{S_y}_{n\mathbf{k}}W_{n\mathbf{k}}$ in the middle three panels, and $\braket{S_z}_{n\mathbf{k}}W_{n\mathbf{k}}$ in the bottom three panels. Note that the color scales are much narrower for $\braket{S_y}$ and $\braket{S_z}$ than $\braket{S_x}$ in the plots.

As we discussed above, the supercell obscures the identification of the spin-split metal Shockley surface states, due to the dense set of bands near $E_F$ from band folding. Since our focus is the Shockley surface states with well-defined characters in the unit cell, we employ the band unfolding technique in Sec. \ref{sec:unfolding} to reveal the unit-cell character of the interface Bloch states. To be specific, when $W_{n\mathbf{k}}$ from Eq. \eqref{eq:weight} is large, the corresponding supercell state $\ket{n\mathbf{k}}$ has strong unit-cell character with crystal momentum $\mathbf{k}$. Therefore, instead of coloring the band structure using $\braket{S_x}$, $\braket{S_y}$, and $\braket{S_z}$ directly, we multiply them by $W_{n\mathbf{k}}$ in Fig. \ref{fig3:7H-Au-MGM}, which facilitates the identification of the dispersive surface states.

The results clearly highlight the dispersion of the Shockley surface states, showing the strength of this choice of color code. For the clean Au (111) surface, the colored bands in Fig. \ref{fig3:7H-Au-MGM}(b) qualitatively agree with the unit-cell band structure [Fig. \ref{fig1:cartoon}(b)], validating our approach. Importantly, the $\braket{S_y}$ and $\braket{S_z}$ are nearly zero, consistent with the expected unit-cell spin texture along the $k_y$ axis [Fig. \ref{fig1:cartoon}(d)].

Comparison between the clean Au (111) surface and the chiral molecule--metal interfaces reveals molecular effects in modulating the metal surface states. As immediately observed from Fig. \ref{fig3:7H-Au-MGM}, the interface bands feature small but finite $\braket{S_y}$ and $\braket{S_z}$ values, along the ($-$M)--$\Gamma$--M line. This indicates that the molecular adsorption rotates the spin off the $x$ axis, an effect we will carefully examine in Sec. \ref{sec:texture} below. The extra non-parabolic features in Fig. \ref{fig3:7H-Au-MGM}(d)(f)(g)(i) reflect the interplay between metal band folding and molecule--metal orbital hybridization. Within the energy window ($E_F\pm 0.5$ eV) of Fig. \ref{fig3:7H-Au-MGM}, we have checked that all states are largely localized on the metal. The highest occupied molecular orbital is about 1.09 eV below $E_F$, while the lowest unoccupied molecular orbital is about 1.38 eV above $E_F$. So the effect reflects the perturbation of metal states by mixing in some molecular components, rather than strong molecular character.

\begin{figure}[!ht]
\centering
\includegraphics[width=\columnwidth]{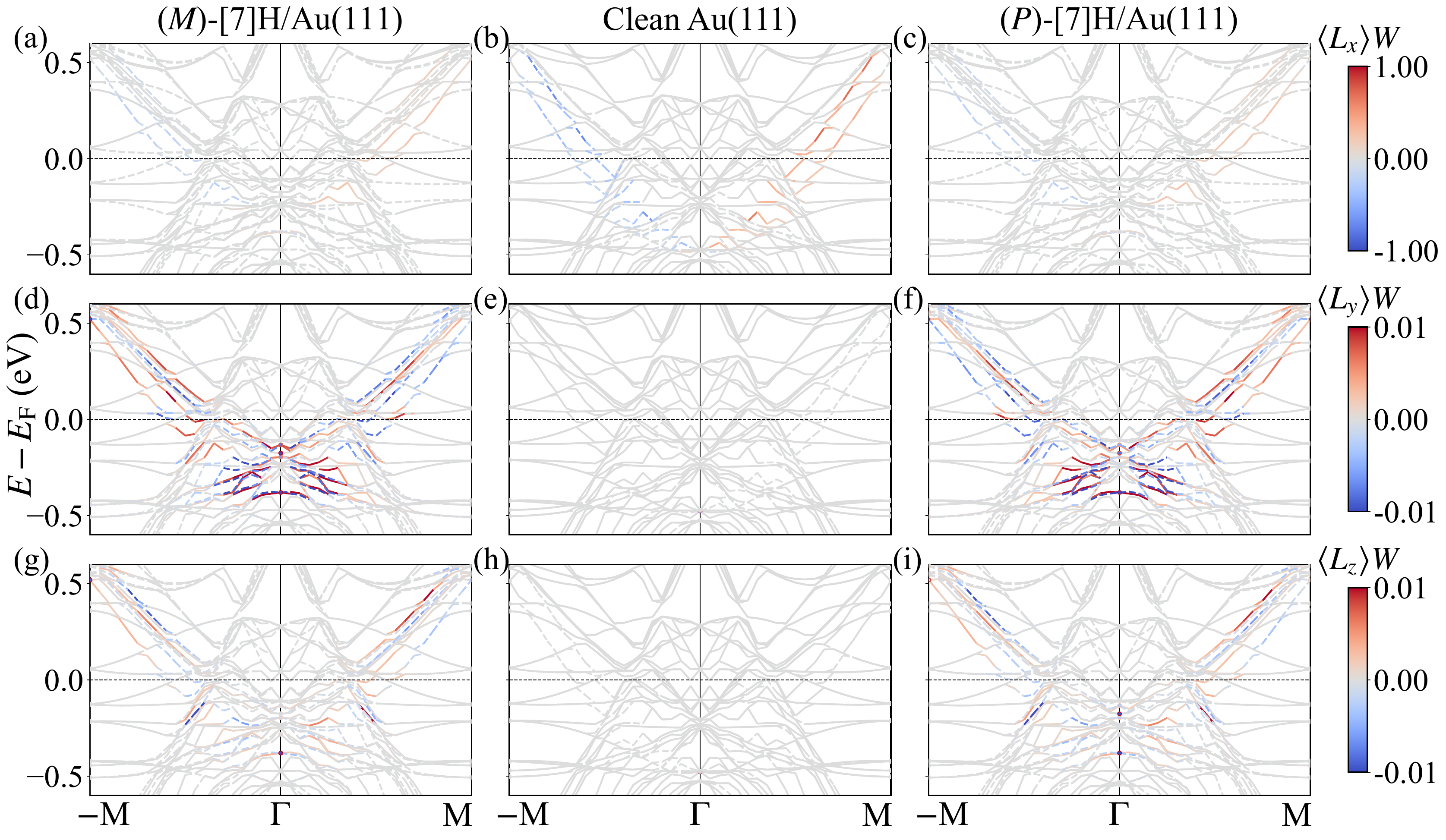}
\caption{Similar to Fig. \ref{fig3:7H-Au-MGM}, but the bands are color-coded by the OAM expectation values [Eqs. \eqref{eq:lz} and \eqref{eq:lxy}] multiplied by the unit-cell character [Eq. \eqref{eq:weight}]: (a)(b)(c) $\braket{L_x}_{n\mathbf{k}}W_{n\mathbf{k}}$, (d)(e)(f) $\braket{L_y}_{n\mathbf{k}}W_{n\mathbf{k}}$, and (g)(h)(i) $\braket{L_z}_{n\mathbf{k}}W_{n\mathbf{k}}$. Note that the color scales are much narrower for $\braket{L_y}$ and $\braket{L_z}$.}
\label{fig4:7H-Au-OAM}
\end{figure}

Fig. \ref{fig4:7H-Au-OAM} colors each state $\ket{n\mathbf{k}}$ in the electronic band structure using OAM expectation values $\braket{L_x}_{n\mathbf{k}}$, $\braket{L_y}_{n\mathbf{k}}$, and $\braket{L_z}_{n\mathbf{k}}$ [Eqs. \eqref{eq:lz} and \eqref{eq:lxy}], multiplied by the unit-cell character $W_{n\mathbf{k}}$ determined from Eq. \eqref{eq:weight}, in a similar manner as Fig. \ref{fig3:7H-Au-MGM}. Again, the clean Au (111) result computed in a supercell matches that of the unit cell from Fig. \ref{fig1:cartoon}, highlighting the Shockley surface states and providing a consistency check. The chiral molecular adsorption-induced changes of the $\braket{L}$ are also qualitatively similar to those of the $\braket{S}$. 

Among all the $\mathbf{k}$ points we sampled, for clean Au (111), the largest $\left|\braket{\mathbf{S}}\right|$ is $0.50\hbar$, while the largest $\left|\braket{\mathbf{L}}\right|$ is $0.74\hbar$. The spin value is consistent with that of the 25-layer Au (111) unit cell. However, we note that the largest $\left|\braket{\mathbf{L}}\right|$ is much smaller than that of the 25-layer Au (111) unit cell reported in Sec. \ref{sec:auuc}, while it is consistent with the result of a four-layer unit cell. Therefore, although the energies of the surface bands at $\Gamma$ do converge for a four-layer slab as we discussed earlier, the $\braket{\mathbf{L}}$ values converge much more slowly with respect to the number of Au layers. Because we model molecule--Au interfaces and a clean Au (111) supercell using the same number of Au layers, we believe that our results provide qualitative, although not strictly quantitative, insight into the molecular modulation of spin and OAM textures. The symmetry relations connecting the two enantiomeric interfaces are discussed together with their momentum-resolved textures in Sec. \ref{sec:texture} below.

\subsection{Spin and OAM textures at molecule--metal interfaces}
\label{sec:texture}

Having examined the band structure along the $k_y$ axis, we now further investigate the spin and OAM textures around the $\Gamma$ point at $E_F$ for the molecule--metal interfaces.

\begin{figure}[!ht]
\centering
\includegraphics[width=\columnwidth]{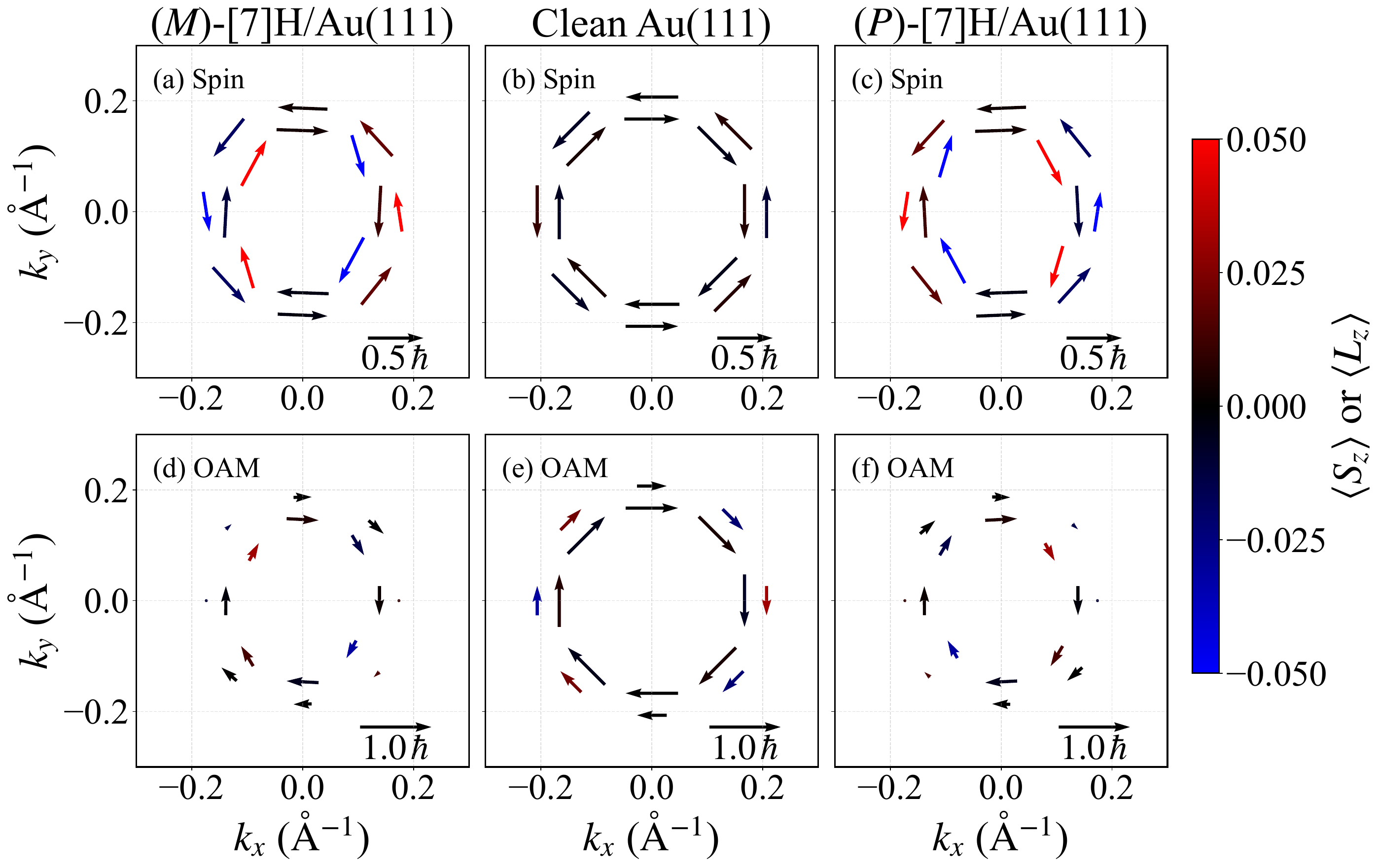}
\caption{(a)(b)(c) Spin and (d)(e)(f) OAM textures near the $\Gamma$ point at $E_F$, for (a)(d) ($M$)-[7]H/Au(111) interface, (b)(e) clean Au (111) surface, and (c)(f) ($P$)-[7]H/Au(111) interface, highlighting the Rashba-split Au surface states. The out-of-plane components, $\braket{S_z}$ and $\braket{L_z}$, are denoted by color.}
\label{fig5:texture}
\end{figure}

Fig. \ref{fig5:texture} compares the textures of the two [7]H/Au(111) interfaces with those of the clean Au (111), for Au-derived surface states within $E_F\pm 5$ meV. Here, we plot $\braket{S}$ and $\braket{L}$ without the unit-cell weight $W$, unlike the weighted plots in the previous section. The clean-surface textures reproduce the qualitative features of the 25-layer results in Fig. \ref{fig1:cartoon}(d)(e). Their OAM magnitudes, however, retain the slab-thickness dependence discussed above.

Adsorption of either [7]H enantiomer changes both the magnitudes and orientations of the spin and OAM vectors. Their in-plane directions deviate from the approximately tangential alignment on clean Au (111), and their out-of-plane components also change. For example, along $k_x$, the clean-surface spin lies predominantly along $y$, whereas the interface states acquire appreciable $\braket{S_x}$ and $\braket{S_z}$. More generally, the sampled out-of-plane spin components are all negligible on the clean Au (111) surface, while they become finite at the [7]H/Au interfaces. 

Importantly, since the ($P$)-[7]H/Au(111) interface was constructed as the mirror image of the ($M$)-[7]H/Au(111) interface through the transformation $y \to -y$, the electronic structure of these two chiral interfaces is connected by symmetry. Under the above mirror operation, the crystal momentum transforms as $(k_x,k_y,k_z)\to(k_x,-k_y,k_z)$, while spin, being an axial vector, transforms as $(\braket{S_x},\braket{S_y},\braket{S_z})\to (-\braket{S_x},\braket{S_y},-\braket{S_z})$. Therefore, between the mirror-related ($M$)-[7]H/Au(111) and ($P$)-[7]H/Au(111), we have
\begin{subequations}
\begin{align}
\braket{S_x}^{\rm P}(k_x,k_y)&=-\braket{S_x}^{\rm M}(k_x,-k_y); \\
\braket{S_y}^{\rm P}(k_x,k_y)&=\braket{S_y}^{\rm M}(k_x,-k_y); \\
\braket{S_z}^{\rm P}(k_x,k_y)&=-\braket{S_z}^{\rm M}(k_x,-k_y).
\end{align}
\label{eq:kxky}
\end{subequations}

For nonmagnetic systems, which is the case studied here, time-reversal symmetry additionally reverses spin between states at $\mathbf{k}$ and $-\mathbf{k}$, i.e.,
\begin{equation}
\braket{S_\alpha}(\mathbf{k})=-\braket{S_\alpha}(-\mathbf{k}),
\label{eq:tr}
\end{equation}
with $\alpha=x,y,$ and $z$. Fig. \ref{fig5:texture} confirms these relations in Eqs. \eqref{eq:kxky} and \eqref{eq:tr}.

Combining Eqs. \eqref{eq:kxky} and \eqref{eq:tr}, as a special case, on the $k_x=0$ axis (the results presented in Sec. \ref{sec:7hau}), we have
\begin{subequations}
\begin{align}
\braket{S_x}^{\rm P}(0,k_y)&=-\braket{S_x}^{\rm M}(0,-k_y)=\braket{S_x}^{\rm M}(0,k_y); \\
\braket{S_y}^{\rm P}(0,k_y)&=\braket{S_y}^{\rm M}(0,-k_y)=-\braket{S_y}^{\rm M}(0,k_y); \\
\braket{S_z}^{\rm P}(0,k_y)&=-\braket{S_z}^{\rm M}(0,-k_y)=\braket{S_z}^{\rm M}(0,k_y),
\end{align}
\label{eq:samek}
\end{subequations}
for non-degenerate states, which applies to the surface states considered in this work, excluding the $\Gamma$ point where Kramers degeneracy occurs. In fact, comparison of the left three panels and the right three panels of Fig. \ref{fig3:7H-Au-MGM} confirms these relations. To be specific, Eq. \eqref{eq:samek} not only relates ($P$)-[7]H/Au to ($M$)-[7]H/Au at reflected momenta on the $k_y$ axis, but also relates the two enantiomeric interfaces at the \emph{same} $\mathbf{k}$ point on the $k_y$ axis.

The OAM textures obey the same symmetry relations. In particular, under the mirror operation $y\to -y$, OAM transforms as $(\braket{L_x},\braket{L_y},\braket{L_z})\to (-\braket{L_x},\braket{L_y},-\braket{L_z})$ as an axial vector, similar to spin. Eqs. \eqref{eq:kxky}-\eqref{eq:samek} apply equally to OAM upon replacing $S_\alpha$ with $L_\alpha$. 

For either [7]H/Au(111) interface, the largest sampled spin and OAM magnitudes are $0.47\hbar$ and $0.45\hbar$, respectively, compared with $0.50\hbar$ and $0.74\hbar$ for the four-layer clean surface. Thus, within this model, adsorption reduces the maximum OAM magnitude substantially more than the maximum spin magnitude. To examine the accompanying orbital redistribution, we transform the complex spherical-harmonic channels in Eq. \eqref{eq:lmlexp} to real atomic-orbital channels. The resulting projections show increased Au $d_{xy}$ and $d_{x^2-y^2}$ weights, decreased $d_{yz}$ and $d_{xz}$ weights, and little change in $d_{z^2}$. These changes indicate adsorption-induced modulation of the orbital composition of the Au-derived surface states. Molecular admixture contributes $s$ and $p$ components, partially reducing OAM magnitudes. Furthermore, the redistributed orbital weight within the Au $d$ manifold discussed above, together with the relative phases and coherences between different orbital components, contributes to the reduced OAM. 

\begin{figure}[!ht]
\centering
\includegraphics[width=\columnwidth]{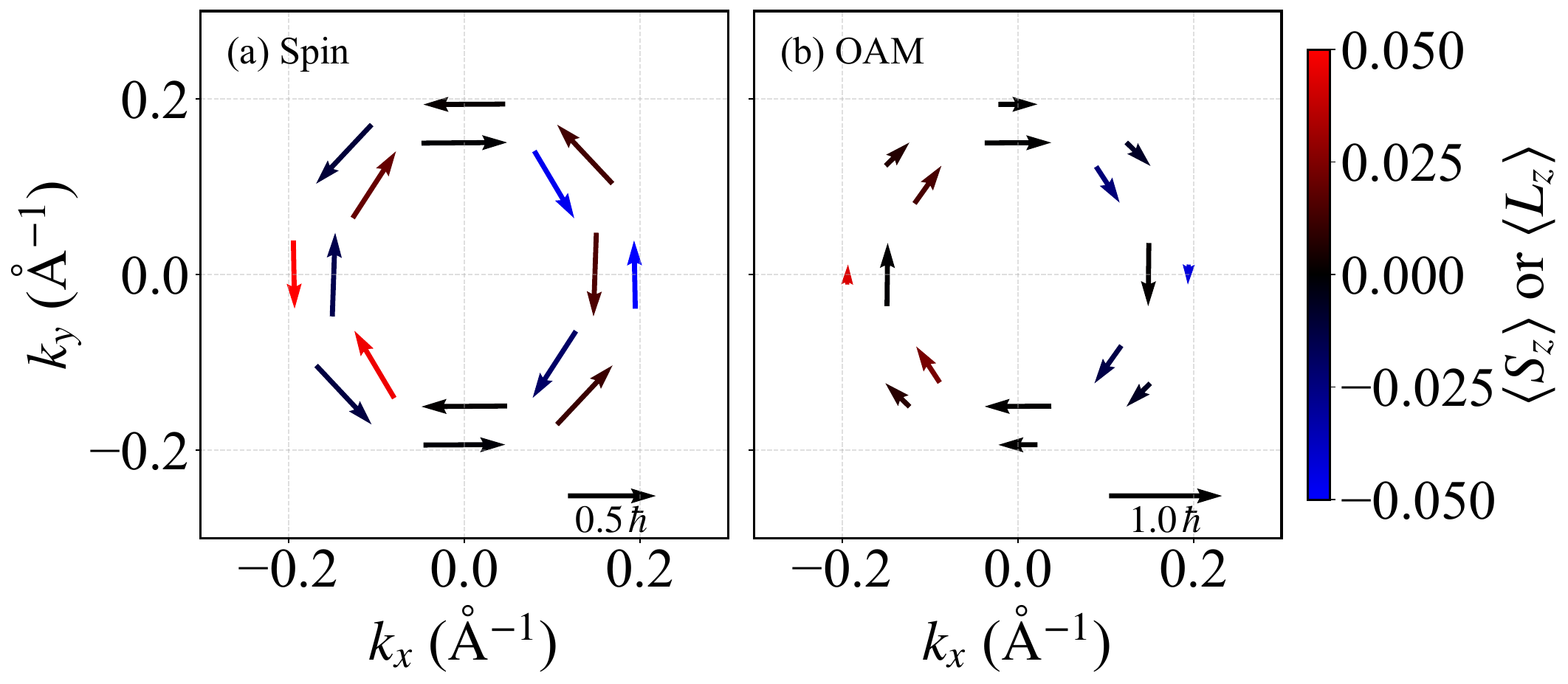}
\caption{(a) Spin and (b) OAM texture near the $\Gamma$ point at $E_F$ for the coronene/Au(111) interface. The out-of-plane components, $\braket{S_z}$ and $\braket{L_z}$, are denoted by color.}
\label{fig6:coronene}
\end{figure}

To identify changes that also occur without molecular chirality, we consider coronene/Au(111), shown in Fig. \ref{fig2:struc}(c). Fig. \ref{fig6:coronene} presents its spin and OAM textures for Au-derived surface states within $E_F\pm5$ meV. As for [7]H adsorption, the spin vectors acquire out-of-plane components. Their presence alone is therefore not a signature of molecular handedness. The largest sampled spin and OAM magnitudes are $0.49\hbar$ and $0.59\hbar$, respectively. Both lie between the clean-surface and [7]H/Au(111) values, with a smaller reduction in the OAM maximum than for [7]H adsorption.

The in-plane vector directions in Fig. \ref{fig6:coronene} remain closer to the clean-surface pattern than those of the [7]H/Au(111) interfaces, while the OAM magnitude and out-of-plane components change. Coronene therefore illustrates adsorption-induced changes that do not require a chiral molecule, whereas the mirror-related [7]H interfaces establish the transformation of the textures upon reflection of the complete interface.

\section{Conclusions}
\label{sec:conclude}

In this work, we have developed a first-principles computational framework, combining DFT calculations with SOC, band unfolding, and an atom-projected analysis to examine how molecular adsorption modifies the spin and OAM textures at chiral molecule--Au(111) interfaces. On clean Au (111), the two Rashba-split Fermi contours have opposite spin helicities but the same sense of OAM circulation. Adsorption modifies the vector orientations and reduces the sampled OAM magnitudes. The two mirror-related [7]H/Au interfaces obey the axial-vector transformation rules together with time-reversal symmetry. The achiral coronene also produces out-of-plane spin components, demonstrating that these components alone do not identify an effect of molecular handedness.

These results motivate momentum-resolved tests of chirality-dependent photoemission. As we discussed above, reflection through the $xz$ plane maps the ($M$)-[7]H/Au interface at $(k_x,k_y)$ onto the ($P$)-[7]H/Au interface at $(k_x,-k_y)$. Therefore, the photoelectron spin polarization measured for ($M$)-[7]H/Au at $(k_x,k_y)$ is symmetry-related to that measured for ($P$)-[7]H/Au at $(k_x,-k_y)$, with the light polarization and detection geometry transformed accordingly. However, importantly, at a generic in-plane momentum $(k_x,k_y)$ away from the symmetry axes, this symmetry relation does \emph{not} fix the difference in photoelectron spin polarization measured for the two enantiomers at the \emph{same} $(k_x,k_y)$ under identical illumination. Angle-resolved measurements around the surface-state Fermi contours could therefore reveal enantiomer-dependent contrast that would have been reduced or obscured by angular averaging. The calculated equilibrium spin and OAM textures characterize the initial states in photoemission and identify where such a comparison between the two enantiomers could be informative, while predicting the actual spin polarization value also requires the photoemission matrix elements and the description of the final states.

\section{Acknowledgments}
We thank Binghai Yan for helpful discussions. This work is supported by an NSF CAREER award, DMR-2044552, and used Bridges-2 at the Pittsburgh Supercomputing Center through Allocation No. PHY220043 from the Advanced Cyberinfrastructure Coordination Ecosystem: Services \& Support (ACCESS) program, which is supported by NSF Grants No. 2138259, No. 2138286, No. 2138307, No. 2137603, and No. 2138296. A.A. acknowledges a Rumble Fellowship and an A. Paul and Carole C. Schaap Endowed Distinguished Graduate Award from the Graduate School at Wayne State University. Z.-F.L. acknowledges an Alfred P. Sloan Research Fellowship, No. FG-2024-21750.

\section{Conflict of Interest Statement}
The authors have no conflicts to disclose.

\section{Author Contributions}
Zhen-Fei Liu conceived the project, designed the computational workflow, supervised the research, and contributed to data analysis. Gyanu Kafle and Amos Afugu performed the calculations and analyzed the data. All authors contributed to writing and revising the manuscript.

\section{Data Availability Statement}
The analyzed data that support the findings of this study are available within the article. The raw numerical data are available from the corresponding author upon reasonable request.

\bibliography{lib}

\end{document}